\documentstyle[pra,aps]{revtex}
\begin{document}
\draft
\title{De Broglie wave, spontaneous emission and
 Planck's radiation law according to stochastic
 electrodynamics}

\maketitle \begin{center}
  O. A. Senatchin
\end{center}

\begin{center}
 Institute of Theoretical and Applied Mechanics,
\end{center}
\begin{center}
 SB Academy of Sciences, Institutskaia
St. 4/1, Novosibirsk, Russia 630090
\end{center}
\begin{center}
 E-mail: olsenat@itam.nsc.ru ~ and ~ olsenat@mail.ru
\end{center}

\date{11 January 2001}
\begin{abstract}

The idea about a quantum nature of Planck's blackbody radiation
law is deeply rooted in minds of most physicists. Einstein's
work, in which the coefficients of spontaneous and induced
emission were introduced, has always been regarded as a proof
that quantum energy discreteness of an atom plays a crucial role in
the derivation of this law. In our paper we avoid this standpoint.
It may be shown  that the de Broglie wave assigned to every material
particle is a result of interaction of the particle with zero-point
vibrations of electromagnetic ground field. The energetic spectrum of
a harmonic oscillator is obtained from this fact within classical
physics which coincides with the quantum result. Thus, it is explained here
how the energy discreteness came into existence in stochastic
electrodynamics (SED) --- the classical electrodynamics with classical
electromagnetic zero-point radiation. Next we reconsider the Einstein
work from the viewpoint of SED and derive the Planck formula.
\end{abstract}

\section{Introduction}
\label{sec:level1}

Every time when we discuss the Planck radiation law, the
quantum radiation theory arises in its monumentality before our
mental vision. Its method of derivation of the probability
coefficients for spontaneous and induced emission practically has not
changed since the appearance of Dirac's fundamental
work [1] in 1927. But is this method so flawless? Does an eye
not catch any contradictions that can be resolved from the
positions of modern physics? As early as 1939  V. L. Ginzburg [2] paid his
attention to one of such contradictions: an absence of correct limiting
transition from quantum theory to
classical theory at $\hbar \rightarrow 0$. In fact, in quantum
theory the intensity of spontaneous emission
\begin{displaymath}
W_{mn}^{sp}=A_{mn} \hbar \omega_{mn}
\end{displaymath}
is equal to zero at the limit $\hbar \rightarrow 0$. At the same
time, the zero-point fluctuations go to zero. But in
traditional classical theory the spontaneous emission  \em is present, \em
in spite of the zero-point radiation absence. The second
contradiction is also obvious. If the existence of the zero-point
radiation stems naturally from the bases of quantum theory, why
it is absent in the final density radiation expression? Thus,
for the derivation of Planck's formula after quantization of
electromagnetic field and obtaining the coefficients $A_{mn}$
and $B_{mn}$ one must refer to the Einstein method [3] of 1917.
However, only the expression

\begin{equation}
\rho \left( {\omega ,T} \right) = \frac{{\omega ^2 }}{{\pi ^2 c^3
}}\frac{{\hbar \omega }}{{e^{{{\hbar \omega } \mathord{\left/
 {\vphantom {{\hbar \omega } {kT}}} \right.
 \kern-\nulldelimiterspace} {kT}}}  - 1}}
\end{equation}
may be derived this way. The zero-point term $(\omega ^2 /\pi ^2 c^3)(
\hbar \omega /2)$ is absent here, but it produces spontaneous
transitions, according to quantum interpretation. Therefore, at the absolute
zero of temperature, the spontaneous emission must be absent also.

In this paper we try to elucidate these questions and, at the
same time, propose a new method of derivation of  Planck's
radiation law. Here we operate with classical ideas only and completely
avoid quantum hypotheses. Such an opportunity is given to us
by stochastic electrodynamics (SED). It is a classical theory with
classical electromagnetic zero-point radiation which playing a role of
the vacuum. SED has had almost 40 years history. It explains quantum
phenomena by methods of classical physics. The most comprehensive recent
surveys on the subject are [4] and [5] ( see also now classical works
[6] and [7] ).

The basic postulate of SED is in the presence of classical
stochastic electromagnetic zero-point (i.e. existing at the
absolute zero of temperature) radiation in the universe. It is
homogeneous, isotropic, Lorentz-invariant, and has the intensity
$\frac{1}{2}\hbar\omega$ per normal mode. According to SED, all
quantum effects, by their essence, are the interaction effects
of matter with the zero-point radiation. Indeed, there have been
made many calculations in SED which prove complete connections
with the results of quantum mechanics (QM) on van der Waals forces
[8-10], Casimir's effect [11,12], correct behavior of a simple
harmonic oscillator [13,14]. The atomic stability [6,15],
the electronic diamagnetism of metals [16], the thermal effects
of acceleration [17,18] have been explained. There are works
concerning classical hydrogen atom [19] which, as yet, have not
reached the desirable goal. But in our work, for the first time,
have obtained the exited states of harmonic oscillator,
and the result, we expect, opens a new avenue of attack on the
hydrogen atom problem.

The electromagnetic theory of gravitation [20] was proposed
by H. E. Puthoff in 1989 on the ground of one old idea of A. D. Sakharov.
It has initiated the investigations of B. Haisch and co-workers [21-24]
on inertia. Actually, very interesting question arises: can one
change the physical, inertial mass of an object situated in a
container by changing the density of electromagnetic field enclosed
in it? The calculations made in works [21] and [23], even though
are too complicated and overweighed by details that hampers to see
the physical essence of the subject, give very promising indications
that it is true. This result is obtained in the present time only by methods
of SED.

The derivation of Planck's radiation law within  classical physics
was put forward long ago, mostly in the works of T. H. Boyer
[25-29]. However, only one method [29] was accepted as a correct one.
We believe that our method is also correct and it has a close
connection with the Einstein-Hopf model, regarded by T. H. Boyer in
work [25]. In one of our future papers we are hoping to tie up our main
equation (22) with the equation (35) of the T. H. Boyer work, proving
by it that the atomic energy discreteness is not the necessary
condition for derivation of Planck's radiation law.

Following the proponents of SED, we take seriously the presence of
homogeneous isotropic Lorentz-invariant electromagnetic  field
with the density radiation
\begin{equation}
\rho _{zp}  = \frac{{\hbar \omega ^3 }}{{2\pi ^2 c^3 }}
\end{equation}
in the universe at the absolute zero of temperature. Therefore,
we expect the following form of  Planck's radiation law
\begin{equation}
\rho \left( {\omega ,T} \right) = \frac{{\omega ^2 }}{{\pi ^2 c^3
}}\left( {\frac{{\hbar \omega }}{{e^{{{\hbar \omega }
\mathord{\left/ {\vphantom {{\hbar \omega } {kT}}} \right.
\kern-\nulldelimiterspace} {kT}}}  - 1}} + \frac{\hbar \omega }{2}  } \right).
\end{equation}

In order to achieve this goal, our paper is arranged as follows.
In the derivation of Eq. (3) we lean upon  Einstein's paper
[3] of 1917. We shortly remind its content in Sec.\
\ref{sec:level2}. It is the irony of fate that in this paper
Einstein tried to show a reality of quanta. However, such an
interpretation of absorption and emission processes of the
two-level system is far from being necessary. In Sections
\ \ref{sec:level3} and \ref{sec:level4} we show that this system may be
classical, in fact, by supplying the Einstein reasoning by an idea
of the classical zero-point radiation. The proof of our main
result is made in Sec.\ \ref{sec:level5}. Most part of calculations
is placed in Appendix.

\section{Einstein's derivation}
\label{sec:level2}

Let us recall, at first, the Einstein fundamental work [3].
More contracted exposition of its ideas was presented in his with
Ehrenfest work [30]. Allow us to repeat their considerations.

An atom can be found in two states $Z$ and $Z^*$ with energies
$\varepsilon$ and $\varepsilon^*$, respectively,
($\varepsilon^*>\varepsilon$). Assume their difference be equal
$\varepsilon^*-\varepsilon=\hbar\omega$. The atom is immersed in
radiation of density $\rho$ and is in thermal equilibrium with
it. Therefore, the level populations obey the canonical
distribution law
\begin{equation}
n \sim e^{ - {\varepsilon  \mathord{\left/ {\vphantom
{\varepsilon  k}} \right. \kern-\nulldelimiterspace} k}T} ,n^ *
\sim e^{ - {{\varepsilon ^ * } \mathord{\left/ {\vphantom
{{\varepsilon ^ *  } {kT}}} \right. \kern-\nulldelimiterspace}
{kT}}} ,\frac{{n^ *  }}{n}\sim e^{ - {{\hbar \omega }
\mathord{\left/ {\vphantom {{\hbar \omega } {kT}}} \right.
\kern-\nulldelimiterspace} {kT}}}.
\end{equation}

For transitions of the atom from $Z^*$-state to $Z$-state
during the time interval $dt$, we have the probability
\begin{equation}
dw^*=(a^*+b^*\rho)dt,
\end{equation}
where $a^*$ is the spontaneous emission coefficient, $b^*$ is the
induced emission coefficient. For $Z \rightarrow Z^*$ atom
transitions, we have the probability
\begin{equation}
dw=b\rho dt.
\end{equation}
Hence, in equilibrium, the following equation is true
\begin{equation}
n^*(a^*+b^*\rho)=nb\rho.
\end{equation}
Using the Eq. (4) and the fact that at $T \rightarrow \infty$,
$\rho \rightarrow \infty$ and, therefore, $b=b^*$, we go to
\begin{equation}
\rho  = \frac{{{{a^ *  } \mathord{\left/ {\vphantom {{a^ *  } {b^
*  }}} \right. \kern-\nulldelimiterspace} {b^ *  }}}}{{e^{{{\hbar
\omega } \mathord{\left/ {\vphantom {{\hbar \omega } {kT}}}
\right. \kern-\nulldelimiterspace} {kT}}}  - 1}} .
\end{equation}

The relation $a^*/b^*$ can be obtained by comparison of
Eqs. (8) and (1). Obviously, if Einstein had had the correct
microscopic theory, he could have obtained the $a^*$ and
$b^*$ coefficients immediately from it. In Sec. \ref{sec:level3}
we will propose such a microscopic theory which
is alternative to Dirac's quantum theory of radiation. More
specifically, it already exists under the name of ``stochastic
electrodynamics'', and we extend it by using the Einstein
work, that has been exposed in this section.

\section{Qualitative introduction to stochastic
electrodynamics}
\label{sec:level3}

Of course, the standpoint of SED is extremely attractive for those who have
strong inner protest against the absence of space-time description in the
atomic world, against the duality of physics. Let us
consider qualitatively some phenomena regarded earlier as particular
quantum.

{\bf Photoeffect}. Conductivity electrons in metal interact
with the zero-point radiation. Hence, they have the kinetic
energy $\frac{1}{2}\hbar\omega$ on a frequency $\omega$. Incoming
radiation is involved in a resonance process with the zero-point
vibrations on a frequency $\omega$, and the emitted electrons
have the energy proportional to $\hbar\omega$. Thus, the energy
of photoelectrons depends only on the frequency, not on the
intensity of incoming radiation.

{\bf Stability of an atom in the ground state}. Why doesn't
the radiating electron fall on the nucleus? Because it absorbs
energy continuously from the zero-point radiation. The balance of energy
between  emitting energy of the electron rotating
around the nucleus and its absorbing energy dictates the radius of
the principal orbit.

{\bf The existence of stationary atomic states}. Let an
electron move around the nucleus on the circular orbit with
some frequency $\omega_0$. Such a motion may be replaced by
harmonic vibrations of two, perpendicular to each other, linear
oscillators of natural frequency $\omega_0$ and phases shifted
on $\pi/2$. The oscillators interact with the zero-point
radiation within a short frequency interval near $\omega_0$. It
would be natural to suppose that the stationary states (the
ground state, in any case) are formed under an energetic balance  of
the electron and the field. Louis de Broglie assigned the
wavelength to any material particle  by extending relations
$E=\hbar\omega$ and $\vec p = \hbar \vec k$ that had been
applicable to the light only before. From the standpoint of SED the
physical meaning of this procedure becomes clear enough. The
energy and momentum of a particle (two oscillators) are
equalled to the mean energy and to the mean momentum of
fluctuations of the interacting field. An atom emits
narrow lines of a spectrum, because an electron may rotate
stationary only on the orbits which include an integer number of
de Broglie waves. On the other orbits the de Broglie wave would
cease because of phases differences.

{\bf One and two-slit experiments}. Let us imagine a wall
with a slit and the zero-point radiation around it. The zero-point
radiation can be essentially homogeneous and isotropic only far
from matter. Near matter, the pattern of electromagnetic
zero-point radiation is modified to reflect presence of the slit.
If an electron or other particle goes through the slit, it will
be influenced by the field fluctuations. The probability of its
deflection, when it achieves the far screen, will be accounted by
the de Broglie wave diffraction.

If a wall has two slits, the zero-point radiation pattern will
reflect this fact. It is evident that every particular electron
goes through only one of the two slits, and the diffraction
picture arises as a consequence of the wavelike properties of
the zero-point radiation.

\section{Derivation of energy spectrum for harmonic oscillator}
\label{sec:level4}

Let us look at the two-level model of Einstein from the new point
of view. Now, a discreteness of stationary states can be
understood qualitatively. In this Section we will show quantitatively,
how the existence of the ground and exited states are explained
in SED for Hertz's resonator, extending the properties of which,
Einstein came to his model.

In SED, just as in quantum electrodynamics (QED), a
harmonic oscillator is the most favorite object. At a time unit
interval it emits the energy (see Appendix Eq. (A4)):
\begin{equation}
W_{em}  = \frac{1}{3}\frac{{e^2 \omega _0^4 r_0^2 }}{{c^3 }}.
\end{equation}
At the same time it absorbs the energy (see Appendix Eq. (A19)):
\begin{equation}
W_{abs}  = \frac{1}{3}\frac{{e^2 \hbar \omega _0^3 }}{{mc^3 }}
\end{equation}
from the zero-point radiation.
Equating (9) and (10), we obtain
\begin{equation}
m \omega_0 r^2_0=\hbar
\end{equation}
which is the Bohr condition for atomic ground state. In the same manner
formula (11) was first obtained in the paper [15], where H.E.Puthoff has corrected
the arithmetic error 3/4 made by T.H.Boyer in the work [6]. His result,
made for two oscillators, is valid also for one. In this case the oscillator
full energy
\begin{equation}
{\rm E} = \frac{{m\omega _0^2 r_0^2 }}{2}
\end{equation}
is equal to
\begin{equation}
\varepsilon=\frac{1}{2}\hbar\omega_0.
\end{equation}

The result can be derived by another way. It is more common and
was already used in the early quantum theory. This is the
introduction of the de Broglie wave. If we put one de Broglie
wave on the principal orbit, then the relation
\begin{equation}
2\pi r_0  = \lambda _0  = \frac{{2\pi \hbar }}{{m\upsilon _0 }}
\end{equation}
will be true, or
\begin{equation}
r_0^2  = \frac{\hbar }{{m\omega _0 }}.
\end{equation}
By substituting Eq. (15) into (12), we obtain again the
expression (13). It is important to remember that the appearance
of the de Broglie wave (14), in our interpretation, connected with
the idea that the particle moving with a constant velocity
extracts from the zero-point radiation  \em a single frequency \em
corresponding to the equality of momenta for the particle
(constituting of two oscillators) and the fluctuations of radiation.

The further extension on the exited states, by such simple  way,
however, does not give the required results. We would obtain
the states with energies $\hbar\omega$, $\frac{3}{2}\hbar\omega$,
$2\hbar\omega$ and so forth, that is inconsistent with the quantum
formula. Thus, let us consider the atom-field interaction more
closely. On this way, in the future, we can refuse from the idea of
the de Broglie wave completely, especially, from the idea of moving
on the orbit the de Broglie wave.

To understand what pattern the radiation has in vicinity of the atom,
let us forget, for a while, the model of rotating around the nucleus
electron and look at the equivalent (in a sense of the wave processes)
picture of two perpendicular to each other oscillators. One is on the
$x$ axis, another on the $y$ axis. Let the monochromatic wave
$\xi _1  = A_0 \cos \omega _0 (t - \frac{x}{c})$ propogate  to positive
direction of $x$ axis; $c$ is a velocity of the wave. It extracts
a single wave with the same frequency and phase from all
vibrations of the zero-point radiation, propogating into the
opposite direction, $\xi _2  = A_0 \cos \omega _0 (t +
\frac{x}{c})$. As a result the standing wave forms
\begin{equation}
\xi  = \xi _1  + \xi _2  = 2A_0 \cos 2\pi \frac{x}{\lambda }\cos
2\pi \frac{t}{T},
\end{equation}
here $\lambda=cT$ and $T=2\pi/\omega_0$. Its amplitude is equal
to zero in knots, when
\begin{displaymath}
\cos 2\pi \frac{x}{\lambda } = 0
\end{displaymath}
or
\begin{displaymath}
2\pi \frac{x}{\lambda } =  \pm (2n + 1)\frac{\pi }{2},\qquad    n=1,2,3,\ldots .
\end{displaymath}
The knots are located on the distances
\begin{equation}
x =  \pm (2n + 1)\frac{\lambda }{4}
\end{equation}
from the origin.

If we return back to the model of the rotating electron and try
to combine it with the resulting field pattern (16), we will go to
the relation $x=r$. In this case, the electron occurs in the
knots of the standing wave with every revolution, and, therefore,
will be able to conserve the uniform motion on the orbit. Because
the equality (13) is valid for
the ground state (when there is only one standing halfwave inside
the orbit), it would be natural to assume that the equation
\begin{equation}
\vec p = \frac{\hbar }{2}\vec k
\end{equation}
is also valid for all states, since the momentum transferred from
the zero-point radiation to the electron is the same in any state.
Or, in a nonrelativistic approximation
\begin{displaymath}
mv\vec e=\frac{\hbar}{2} \frac {2\pi}{\lambda} \vec n.
\end{displaymath}
Geometrical considerations require that $\vec n /\vec e=2/ \pi$,
therefore
\begin{equation}
\lambda  = \frac{{2\hbar }}{{m\omega _0 r}}.
\end{equation}
Setting the formula into Eq. (17), at $x=r$, we go to
\begin{equation}
r^2  = (n + \frac{1}{2})\frac{\hbar }{{m\omega _0 }}.
\end{equation}
As we have two oscillators, then using relation (12) we arrive at
\begin{equation}
E_n  = (n + \frac{1}{2})\hbar \omega _0
\end{equation}
which is in full accordance with the quantum result.

Thus, the idea of the de Broglie wave is appeared to be only a trick
to account the interaction of multiply periodic atomic system with
the zero-point radiation.

It is interesting now to make a historical remark. When, in 1926, Bohr
and Schr\"odinger argued about the interpretation of wave
mechanics, and the latter put the statement that quanta are
unnecessary for understanding the quantum phenomena, Bohr referred
to the derivation of Planck's radiation law in Einstein's paper
[3] as to his basic counterargument (the details of the
discussion were vividly described by Heisenberg in [31] ). We will
show here that such a reference has a fragile basis.

\section{Derivation of Planck's radiation law within a framework of SED}
\label{sec:level5}

Our next step will be to obtain in a framework of SED such an
equation that is a counterpart to Einstein's Eq.~(7). For this
purpose we must, first at all, realize, in a limiting clarity,
what is the spontaneous emission. After the appearance of
Einstein's work [3] and Dirac's theory [1], in a physical
literature frantic and fruitless discussions began [2, 32-34] on
the following questions. Have the spontaneous emission either
quantum or classical origin? Does the zero-point radiation induce
the spontaneous emission? Why does the zero-point radiation affect on
matter twice as much as usual thermal radiation?

We believe that all those discussions occur because a mess of
classical and quantum ideas. In stochastic electrodynamics the
spontaneous emission has the unique and clear physical meaning:
it is the emission of classical system in the absence of external
fields. A classical Hertz's oscillator emits energy (9). This is the
spontaneous emission. It presents both in the exited state and in
the ground state. In the ground state it cancels out by
absorption completely and the atom does not emit, whereas in the
exited state the cancelation is in part only, and the atom emits.
Naturally, the oscillator must absorb energy from the
zero-point radiation in the both states.

It is convinient to exhibit all said in the tabulated form.

\begin{center}
\begin{tabular}{|c|c|c|}
\hline
in Einstein's work &  & in SED \\ \hline
$(n^*a^*+n^*b^*\rho_T)\hbar\omega$ & emitted energy &
$n^*a^*+na+n^*b\rho_T$
\\ \hline
$\hspace*{\fill}n
b\rho_T\hbar\omega$ & absorbed energy &
$nb\rho_0+n^*b^*\rho_0+nb\rho_T$\\ \hline
\end{tabular}
\end{center}

In spite of the fact that the number of terms are doubled in SED,
by comparison with Einstein's work, a striking symmetry is
appeared now that was absent before. A physical meaning of the
coefficients is altered in some details. Now $a$ is the energy
emitted by one atom and $b\rho_0$ is the energy absorbed from the
zero-point radiation by one atom. The asterisk means, as
usual, belonging to the exited state. In SED we do not speak anymore
about a probability of transitions, since we have no quanta
of energy $\hbar\omega$ now. Energy emits and absorbs
continuously. However, taking into account the assumption that
the transition times is much smaller than the living times in the
ground and in the exited state, we can write down the energy
balance equation as
\begin{equation}
n^*a^*+na+n^*b\rho_T=n^*b\rho_0+nb\rho_0+nb\rho_T.
\end{equation}
Here, just like Einstein, we have used the limit
$T\rightarrow\infty$, from which follows that $b^*=b$. It is
not difficult to see from Eq. (22) that in SED, in contrast to
Bohr's theory, the emission processes occur when an electron
is in the exited state, not during a time of transitions between
the states.

We consider now an another limit, $T=0$,
\begin{displaymath}
n^*a^*+na=n^*b\rho_0+nb\rho_0,
\end{displaymath}
then
\begin{displaymath}
n^*(a^*+b\rho_0)=n(-a+b\rho_0),
\end{displaymath}
and we find that
\begin{equation}
a=b\rho_0.
\end{equation}
The result is already known. It means the existence of the full
energy balance in the ground state at $T=0$. The equation (22)
reduces to
\begin{equation}
n^*a^*+n^*b\rho_T=n^*b\rho_0+nb\rho_T.
\end{equation}
Furthermore, since $\varepsilon^*=\frac{3}{2}\hbar\omega_0$, we have
$a^*=3a$ (see Eq. (6A)), therefore
\begin{equation}
n^*(2\rho_0+\rho_T)=n\rho_T.
\end{equation}
From this equation it is easy to obtain the radiation law for the function
\begin{equation}
\rho=\rho_0+\rho_T
\end{equation}
which is a reduced form of Eq. (3). Indeed, we have
\begin{equation}
n^*(\rho_0+\rho)=n(\rho-\rho_0)
\end{equation}
or
\begin{equation}
\rho  = \rho _0 \frac{{e^{{{\hbar \omega } \mathord{\left/
{\vphantom {{\hbar \omega } {kT}}} \right.
\kern-\nulldelimiterspace} {kT}}}  + 1}}{{e^{{{\hbar \omega }
\mathord{\left/ {\vphantom {{\hbar \omega } {kT}}} \right.
\kern-\nulldelimiterspace} {kT}}}  - 1}}.
\end{equation}
Finally, substituting the explicit expression of $\rho_0$ into (28), we
obtain
\begin{displaymath}
\rho  = \frac{{\hbar \omega ^3 }}{{\pi ^2 c^3 }}\left(
{\frac{1}{{e^{{{\hbar \omega } \mathord{\left/ {\vphantom {{\hbar
\omega } {kT}}} \right. \kern-\nulldelimiterspace} {kT}}}  - 1}}
+ \frac{1}{2}} \right).
\end{displaymath}

\section{Conclusion}
\label{sec:level6}

In the textbooks on quantum electrodynamics (QED) and quantum
optics one can often find the statement that the spontaneous
transitions appear as a consequence of atomic electrons
interaction with the zero-point vibrations of electromagnetic
field, or, the photon vacuum, because there can not be
self-emergent transitions in Nature, going on without any
interactions. Needless to say, the latter is true. However, the
cause of the Hertz resonator  emission in classical
electrodynamics is not the interaction with a vacuum, but the
interaction with a self-field, a self-action. In this paper we
adhere strictly to the classical concept. In the case a mess
of ideas does not arise.

According to quantum theory, an atom can have only a discrete set
of states $Z_1,Z_2,\ldots,Z_n$ with eigenenergies
$\varepsilon_1,\varepsilon_2,\ldots,\varepsilon_n$. Einstein,
studying the problem of derivation of Planck's radiation law, has
proposed the model of an atom which have only two states: $Z$
--- the ground state, from where the spontaneous transitions are
absent, and $Z^*$ --- the exited state from where occur as
spontaneous as stimulated transitions. From a view point of
stochastic electrodynamics, classical theory with
classical electromagnetic zero-point radiation, the existence of
the ground state $Z$ is explainable by an energy balance between
self-emission and absorption from the zero-point radiation. If
we assume that the exited state is formed in the similar way (i.e. there
are both absorption and emission for it), then, as we had
shown, it is easy to derive  Planck's radiation law within a
framework of classical physics. Einstein's work on spontaneous
and induced radiation would present itself in another light  before
us. After making a proper interpretation to the spontaneous
emission, we obtained the coefficients $a$ and $b$ without
the postulates of quantum mechanics and the second quantization
procedure. It proves that the zero-point fluctuations are not the
reason of spontaneous transitions. The connection between SED
and QED is in using the same properties of a simple harmonic
oscillator. However, their interpretations of electromagnetic
radiation are diametrically opposite. And today we could incline to
prefer SED because of its simplicity and duality absence.

After completing the calculations, exhibited in this paper,
the author became aware that he is hardly the first who had hit
upon a connection of the de Broglie wave with the zero-point
radiation [35].

\section*{Acknowledgement}

I am indebted to Yu. M. Shelepov for invaluable collaborative
discussion throughout this effort.

\appendix
\section*{}

The motion equation of a harmonic oscillator have a form
\begin{equation}
m\ddot x + \omega _0^2 x = 0.
\end{equation}
Its solution
\begin{equation}
x = r_0 \cos \omega t
\end{equation}
shows that the acceleration
\begin{equation}
\ddot x =  - \omega ^2 r_0 \cos \omega t
\end{equation}
yields to the charged particle emission with the intensity
\begin{equation}
W_{em}  = \frac{{2e^2 }}{{3c^3 }}\left\langle {\ddot x^2 }
\right\rangle  = \frac{{e^2 \omega _0^4 r_0^2 }}{{3c^3 }},
\end{equation}
because
\begin{displaymath}
\left\langle {\cos ^2 \omega t} \right\rangle  = \frac{1}{2}.
\end{displaymath}
Full oscillator energy $E=T+V$ is equal to
\begin{equation}
E = \frac{{m\omega _0^2 x^2 }}{2} + \frac{{mx^2 }}{2} =
\frac{{m\omega _0^2 r_0^2 }}{2}.
\end{equation}
Substituting $r_0$ of (5) into Eq. (4) we go to
\begin{equation}
W_{em}  = \frac{2}{3}\frac{{e^2 \omega _0^2 E}}{{mc^3 }}.
\end{equation}
   The zero-point radiation can be written as a sum over plane waves
\begin{displaymath}
E _{zp}\left( {r,t} \right) =
\sum\limits_{\lambda  = 1}^2 {d^3 k\hat \varepsilon } \left(
{\vec k,\lambda } \right)h\left( \omega  \right) \cos \left[
{\vec k\vec r - \omega t + \theta \left( {\vec k,\lambda }
\right)} \right]
\end{displaymath}
\begin{equation}
=\sum\limits_{\lambda = 1}^2 {\int {d^3 k\hat \varepsilon \left(
{\vec k,\lambda } \right)} } \frac{{h\left( \omega
\right)}}{2}\left[ {a\left( {\vec k,\lambda } \right)e^{i\left( {\vec
k\vec r - \omega t} \right)} + a^ * \left( {\vec k,\lambda }
\right)e^{ - i\left( {\vec k\vec r - \omega t} \right)} } \right] ,
\end{equation}
where $h(\omega)$  is the factor that puts a scale on the wave energy. For the
zero-point radiation the relation
\begin{equation}
h _{zp}^2 \left( \omega  \right) = \frac{{\hbar \omega }}{{2\pi ^2 }}
\end{equation}
is valid,
\begin{displaymath}
a\left( {\vec k,\lambda } \right) = e^{i\theta \left( {\vec
k,\lambda } \right)} ,\quad a^ *  \left( {\vec k,\lambda } \right) =
e^{ - i\theta \left( {\vec k,\lambda } \right)},
\end{displaymath}
$\theta \left( {\vec k,\lambda } \right)$ --- a random
phase  distributed uniformely on the interval $[0,2\pi]$ , independently
distributed for each $\vec k$ and $\lambda$.

To obtain the energy absorbed by the linear harmonic  oscillator from the field
we must solve the linear stochastic equation

\begin{equation}
m\frac{d^2 x}{d t^2} + m\omega _0^2 x - \frac{2}{3}\frac{{e^2 }}{{mc^3 }}
\frac{d^3 x}{d t^3}=
eE_{{\rm }} \left( {0,t} \right)
\end{equation}
Its solution can be found by Fourier method
\begin{equation}
x\left( t \right) = \frac{e}{m}\sum\limits_{\lambda  = 1}^2 \int
d^3 k\hat \varepsilon _x \left( {\vec k,\lambda }
\right)\frac{{h\left( \omega  \right)}}{2}
\left[ {\frac{{a\left( {\vec k,\lambda }
\right)}}{{C\left( \omega  \right)}}e^{i\left( {\vec k\vec r -
\omega t} \right)}  + \frac{{a^ *  \left( {\vec k,\lambda }
\right)}}{{C^ *  \left( \omega  \right)}}e^{ - i\left( {\vec
k\vec r - \omega t} \right)} } \right],
\end{equation}
where
\begin{displaymath}
C\left( \omega  \right) = \omega _0^2  - \omega ^2  - i\Gamma
\omega ^3 \quad\mbox{ and }\quad \Gamma  = \frac{2}{3}\frac{{e^2 }}{{mc^3 }} ,
\end{displaymath}
$\hat \varepsilon \left( {\vec k,\lambda } \right)$ in Eqs. (A7)
and (A9) is a polarization vector. It obeys the relations
\begin{equation}
\hat \varepsilon \left( {\vec k_1 ,\lambda } \right)\hat
\varepsilon \left( {\vec k_2 ,\lambda } \right) = \delta
_{\lambda _1 \lambda _2 },\quad \vec k\hat \varepsilon \left( {\vec
k,\lambda } \right) = 0,
\end{equation}
from those one can obtain the very useful identities for summing over the
two possible polarizations:

\begin{equation}
\sum\limits_{\lambda  = 1}^2 {\varepsilon _i \left( {\vec
k,\lambda } \right)\varepsilon _j \left( {\vec k,\lambda }
\right) = \delta _{ij}  - \frac{{k_i k_j }}{{k^2 }}}.
\end{equation}

Therefore, the absorption energy is
\begin{displaymath}
W  = \int\limits_0^\tau  dt\dot x\left( {eE_{zp}}\right)
= \frac{{e^2 }}{m}\sum\limits_{\lambda _1  = 1}^2
\sum\limits_{\lambda _2  = 1}^2 \int d^3 k_1 \int d^3 k_2 \hat
\varepsilon _{x1} \hat \varepsilon _{x2}{\left(  - i\omega  \right)\frac{{h_1 \left( {\omega _1 }
\right)}}{2}\frac{{h_2 \left( {\omega _2 } \right)}}{2}}
\end{displaymath}
\begin{equation}
\times\left[ {\frac{{a_1 }}{{C_1 }}e^{i\left( {\vec k_1 \vec
r_1  - \omega _1 t} \right)}  - \frac{{a_1^ * }}{{C_1^
*  }}e^{ - i\left( {\vec k_1 \vec r_1  - \omega _1 t} \right)} }
\right] \left[ {a_2 e^{i\left( {\vec k_2 \vec r_2  - \omega _2 t}
\right)}  + a_2^ * e^{ - i\left( {\vec k_2 \vec r_2 - \omega _2
t} \right)} } \right].
\end{equation}
We averaging now on the random phases. The phasis $\theta \left(
{\vec k,x} \right)$ have a normal distribution, i.e. its second
moments are equal to
\begin{displaymath}
\left\langle {\varepsilon ^{i\theta \left( {\vec k_1 ,\lambda _1
} \right)} e^{ - i\theta \left( {\vec k_2 ,\lambda _2 } \right)}
} \right\rangle  = \left\langle {a_1 a_2^ *  } \right\rangle  =
\delta _{\lambda _1 \lambda _2 } \delta ^3 \left( {\vec k_1  -
\vec k_2 } \right)
\end{displaymath}
\begin{equation}
\left\langle {a_1 a_2 } \right\rangle  = 0,\quad \left\langle {a_1^ *
a_2^ *  } \right\rangle  = 0.
\end{equation}
From it follows that
\begin{displaymath}
\left\langle {\left[ {\frac{{a_1 }}{{C_1 }}e^{ - i\omega _1 t}  -
\frac{{a_1^ *  }}{{C_1^ *  }}e^{i\omega _1 t} } \right]\left[
{a_2 e^{ - i\omega _2 t}  + a_2^ *  e^{i\omega _2 t} } \right]}
\right\rangle
\end{displaymath}
\begin{equation}
= \left( {\frac{1}{{C_1 }} - \frac{1}{{C_1^ *  }}} \right)\delta
_{\lambda _1 \lambda _2 } \delta ^3 \left( {\vec k_1  - \vec k_2
} \right) =\frac{{2i\Gamma \omega _1^3 }}{{\left| {C_1 \left( {\omega _1 }
\right)} \right|^2 }}\delta _{\lambda _1 \lambda _2 } \delta ^3
\left( {\vec k_1  - \vec k_2 } \right).
\end{equation}
Thus, averaging, we go to
\begin{equation}
W_{abs}  = \frac{{e^2 }}{m}\sum\limits_{\lambda  = 1}^2 {\int {d^3
k\hat \varepsilon _{1x} \hat \varepsilon _{2x} \frac{{h^2 \left(
\omega  \right)}}{2}\frac{{\Gamma \omega ^4 }}{{\left| {C\left(
\omega  \right)} \right|^2 }}} }.
\end{equation}
Taking into account (A12), the angle integration yields to
\begin{equation}
\sum\limits_{\lambda  = 1}^2 \int d^3 k\hat \varepsilon _{1x}
\hat \varepsilon _{2x}  = \int {d\Omega \left( {1 - \frac{{k_x^2 }}{{k^2 }}}
\right)}\int {dk k^2 }  = \frac{{8\pi }}{3}\int {d\omega \frac{{\omega ^2
}}{{c^3 }}}.
\end{equation}
The function $1/\left| {C\left( \omega  \right)}
\right|^2 $ have a narrow maximum near $\omega=\omega_0$ in the
limit ${{e^2 } \mathord{\left/ {\vphantom {{e^2 } m}} \right.
\kern-\nulldelimiterspace} m} \to 0 $. Therefore, we can invoke the standard
resonance approximation. We change $\omega$ on $\omega_0$  for it everywhere
except the difference $\omega-\omega_0$  and take the integration
limits go to $-\infty$. Then we take the analogous inpact of
$\omega+\omega_0$  into account. So, prove that
\begin{equation}
\mathop {\lim }\limits_{{{e^2 } \mathord{\left/ {\vphantom {{e^2
} m}} \right. \kern-\nulldelimiterspace} m} \to 0} \frac{{e^2
}}{m}\frac{1}{{\left| {C\left( \omega  \right)} \right|^2 }} =
\frac{{3\pi c^3 }}{{4\omega _0^4 }}\left[ {\delta \left( {\omega
- \omega _0 } \right) + \delta \left( {\omega  + \omega _0 }
\right)} \right].
\end{equation}
Indeed
\begin{displaymath}
\frac{e^2 }{m}\int\limits_{ - \infty }^\infty  {d\omega
\frac{1}{{\left( {\omega _0^2  - \omega ^2 } \right)^2 + \Gamma ^2
\omega ^6 }}} = \frac{3}{2}\Gamma c^3 \int\limits_{ - \infty }^\infty  {\left(
\omega  - \omega _0 \right)} \frac{1}{\left( {\omega  - \omega _0}
\right)^2 4\omega _0^2  + \Gamma ^2 \omega ^6 }
\end{displaymath}
\begin{displaymath}
= \frac{3}{4}\frac{{c^3 }}{{\omega _0^4 }}\int\limits_{ - \infty
}^\infty  {d\left( {\omega  - \omega _0 } \right)}
\frac{{{{\Gamma \omega _0^2 } \mathord{\left/ {\vphantom {{\Gamma
\omega _0^2 } 2}} \right. \kern-\nulldelimiterspace} 2}}}{{\left(
{\omega  - \omega _0 } \right)^2  + \left( {{{\Gamma \omega _0^2
} \mathord{\left/ {\vphantom {{\Gamma \omega _0^2 } 2}} \right.
\kern-\nulldelimiterspace} 2}} \right)^2 }} = \frac{{3\pi c^3
}}{{4\omega _0^4 }}.
\end{displaymath}
Substituting (A18) and (A17) into Eq. (A16) we obtain
\begin{equation}
W_{abs}  = \frac{1}{3}\frac{{e^2 \hbar \omega _0^3 }}{{mc^3 }}.
\end{equation}


\begin{references}
\bibitem{1}P. A. M. Dirac, Proc. Roy. Soc. A {\bf 114,} 243 (1927).
\bibitem{2}V. L. Ginzburg, Dokl. Acad. Nauk SSSR, {\bf 23,} 773 (1939);
 {\bf 24,} 130 (1939).
\bibitem{3}A. Einstein, Phys. Zeitschrift {\bf 18,} 121 (1917).
\bibitem{4}L. De la Pe\~{n}a and A. M. Cetto, The Quantum Dice: An Introduction
 to Stochastic Electrodynamics, (Kluwer Acad. Publ., Dordrecht,
 the Netherlands, 1996).
\bibitem{5}D. C. Cole, in Essays on Formal Aspects of Electromagnetic
 Theory, edited by A. Lakhatakia (World Scientific,
 Singapore, pp. 501-532, 1993).
\bibitem{6}T. H. Boyer, Phys. Rev. D {\bf 11,} 790 (1975).
\bibitem{7} L. De la Pe\~{n}a, in Stochastic Processes Applied in Physics and
 Other Related Fields, edited by B. Gomez et al (World Scientific,
 Singapore, 1983).
\bibitem{8}T. W. Marshall, Nuovo Cimento {\bf 38,} 206 (1965).
\bibitem{9}T. H. Boyer, Phys. Rev. A {\bf 7,} 1832 (1973).
\bibitem{10}T. H. Boyer, Phys. Rev. A {\bf 9,} 2078 (1974).
\bibitem{11}T. H. Boyer, Annals of Phys., {\bf 56,} 474 (1970).
\bibitem{12}T. H. Boyer,  Phys. Rev. {\bf 174,} 1631 (1968).
\bibitem{13}T. W. Marshall, Proc. R. Soc. London, A {\bf 276,} 475 (1963).
\bibitem{14}T. H. Boyer, Phys. Rev. D {\bf 11,} 809 (1975).
\bibitem{15}H. E. Puthoff, Phys. Rev. D {\bf 35,} 3266 (1987).
\bibitem{16}T. H. Boyer, Phys. Rev. A {\bf 21,} 66 (1980).
\bibitem{17}T. H. Boyer, Phys. Rev. D {\bf 21,} 2137 (1980).
\bibitem{18}T. H. Boyer, Phys. Rev. D {\bf 29,} 1089 (1984).
\bibitem{19}See, for example,
 P. Claverie and F. Soto, J. Math. Phys. {\bf 23,} 753 (1982).
\bibitem{20}H. E. Puthoff, Phys. Rev. A {\bf 39,} 2333 (1989).
\bibitem{21}B. Haisch, A. Rueda and H. E. Puthoff,
 Phys. Rev. A {\bf 48,} 678 (1994).
\bibitem{22}A. Rueda and B. Haisch, Phys. Lett. A {\bf 240,} 115 (1998).
\bibitem{23}A. Rueda and B. Haisch, Found. Phys. {\bf 28,} 1057 (1998).
\bibitem{24}M. Ibison and B. Haisch, Phys. Rev. A {\bf 54,} 2737 (1996).
\bibitem{25}T. H. Boyer, Phys. Rev. {\bf 182,} 1374 (1969).
 See also the discussion of this paper in
 J. L. Jimenez, L. De la Pe\~{n}a and T. A. Brody,
 Am. J. Phys. {\bf 48,} 840 (1980); T. W. Marshall, Phys. Rev. D
 {\bf 24} 1509 (1981); P. W. Milonni, Phys. Rep. {\bf 25,} 1 (1976).
\bibitem{26}T. H. Boyer, Phys. Rev. {\bf 186,} 1304 (1969).
\bibitem{27}T. H. Boyer, Phys. Rev. D {\bf 27,} 2906 (1983).
\bibitem{28}T. H. Boyer, Phys. Rev. D {\bf 29,} 2418 (1984).
\bibitem{29}T. H. Boyer, Phys. Rev. D {\bf 29,} 1096 (1984).
\bibitem{30}A. Einstein und P. Ehrenfest, Zs. Phys. {\bf 19,} 301 (1923).
\bibitem{31}W. Heisenberg,  Der Teil und das Ganze, M\"unchen, 1969.
\bibitem{32}V. F. Weisskopf, Naturwissenschaften {\bf 27,} 631 (1935).
\bibitem{33}B. Fain,  Nuovo Cimento B {\bf 68,} 73 (1982).
\bibitem{34}V. L. Ginzburg, Usp. Phys. Nauk {\bf 140}, 687(1983) [Sov. Phys.
Usp. {\bf 26}, 713 (1983)].
\bibitem{35}See B. Haisch and A. Rueda, Phys. Lett. A {\bf 268,} 224 (2000),
 and references therein.

\end{references}
\end{document}